\documentstyle[12pt,supercite]{article}

\newcommand{\eps}{\epsilon}
\newcommand{\om}{\omega}

\renewcommand{\sinh}{{\rm sh}\,}

\renewcommand{\coth}{{\rm cth}\,}
\newlength{\digitwidth} 

\def\sqr#1#2{{\vcenter{\hrule height.#2pt
      \hbox{\vrule width.#2pt height#1pt \kern#1pt
        \vrule width.#2pt}
      \hrule height.#2pt}}}

\def\abstracts#1#2#3{{
        \centering{\begin{minipage}{4.62in}\baselineskip=13pt
        \small
        \centerline{\bf Abstract}
        \vspace*{0.2cm}                % W. Janke (July 1, 1992)
        \parindent=0pt #1\par
        \parindent=18pt #2\par
        \parindent=15pt #3
        \end{minipage} }\par}}

\renewcommand{\thefootnote}{\fnsymbol{footnote}}
\begin{document}
\vspace*{-3cm}
\hfill \parbox{4.2cm}{ FUB-HEP 10/94\\
                August 1994 \\ %\today \\
               }\\
\vspace*{1.3cm}
\centerline{\LARGE \bf Test of variational approximation} \\[0.2cm]
\centerline{\LARGE \bf for $\phi^4$ quantum chain} \\[0.2cm]
\centerline{\LARGE \bf by Monte Carlo simulation}\\[0.8cm]
%\centerline{\LARGE \bf by Monte Carlo simulation\footnotemark}\\[0.2cm]
%\footnotetext{xxx}
%\addtocounter{footnote}{-1}
\renewcommand{\thefootnote}{\arabic{footnote}}
\vspace*{0.3cm}
\centerline{\large {\em Wolfhard Janke\/}$^{1}$ 
               and {\em Tilman Sauer\/}$^2$}\\[0.4cm] 
\centerline{\large    $^1$ {\small Institut f\"ur Physik,
                      Johannes Gutenberg-Universit\"at Mainz}}
\centerline{    {\small 55099 Mainz, Germany }}\\[0.15cm]
\centerline{\large    $^2$ {\small Institut f\"{u}r Theoretische Physik,
                      Freie Universit\"{a}t Berlin}}
\centerline{    {\small 14195 Berlin, Germany}}\\[2.50cm]
\vspace*{0.3cm}
\abstracts{}{
We report results of a Monte Carlo simulation of the 
$\phi^4$ quantum chain.
In order to enhance the efficiency of the simulation we 
combine multigrid simulation techniques
with a refined discretization scheme.
The resulting accuracy of our data allows for a 
significant test of an analytical approximation
based on a variational ansatz.
While the variational approximation is well reproduced
for a large range of parameters
we find significant deviations for low temperatures
and large couplings.
}{}
\vspace*{0.5cm}
  \thispagestyle{empty}
      \newpage
       \pagenumbering{arabic}
%
%-------------------------------------------------------------------
       \section{Introduction}
%-------------------------------------------------------------------
%
 
%

The physics of one-dimensional quantum systems has attracted considerable
attention, both experimental and theoretical, for a long time.
Among the methods to treat these systems analytically the 
variational approximation \cite{kleinert_1990,gt,fk86,gtv88a,gtv88b}
has been shown to be a very powerful and useful one \cite{janke}.
Since, however, it is hard to give precise inherent error estimates
for the variational approach it is therefore desirable to check the
method against independently obtained data.
For the $\phi^4$ chain which has been investigated both as a classical
\cite{ss,sazaki,fs93} and as a quantum system
\cite{gtv88b,tm79}
apparently no such independent data exist up to this date.

Precise Monte Carlo simulations of many-particle quantum systems
based on a path-integral representation of the partition function
would provide just such an independent approach for these systems \cite{mc}.
The difficulty here is to achieve sufficient accuracy. Standard path-integral
simulations suffer from well-known draw backs, such as
appreciable systematic errors due to the necessary discretization
and severe slowing down in the continuum limit.
In order to overcome these problems a Fourier Monte Carlo simulation was 
tried some time ago for the closely related sine-Gordon chain
\cite{wr87}.
Even though preliminary data seemed to
reproduce the variational approximation \cite{gtv88a} quite well similar 
results for the $\phi^4$ chain were not obtained.
Unfortunately, a full account of these investigations was never published
\cite{DeRaedtprivate}.
A disadvantage of the method used in Ref.\cite{wr87} is that
it is not based on importance sampling which is a problem 
particularly for unbounded potentials such as the $\phi^4$ double well.

In view of these difficulties it is therefore gratifying that
recently some algorithmic improvements developed for spin systems
and lattice field theories could successfully be transferred to 
path-integral simulations \cite{js92b}.
Multigrid simulation techniques \cite{multigrid} in particular
have been shown to eliminate slowing down in the continuum limit for 
one-particle systems \cite{js93a}. 
It seemed therefore worthwhile to investigate whether these algorithmic
improvements may now render path-integral simulations of quantum chains
sufficiently accurate to allow for a significant comparison with
the variational approximation. 
In this letter we will report simulation data for the $\phi^4$ quantum
chain obtained by combining multigrid simulation techniques with 
a refined discretization scheme. It will be shown that the accuracy of
the data does allow for a qualified judgement about the validity of the 
variational approximation. 

%------------------------------------------------------------------
       \section{The model and variational approximation}
%------------------------------------------------------------------
       \label{sect:4GenDisc}
%------------------------------------------------------------------
The system we are going to discuss is defined by the
partition function
\begin{equation}
    {\cal Z} = e^{-\beta F} = \prod_{i=1}^N\int_{\phi_i(0)=\phi_i(\hbar\beta)}
               {\cal D}[\{\phi_i(u)\}]e^{-{\cal H}[\{\phi_i(u)\}]/\hbar},
 \label{eq:4partitionfunction}
\end{equation}
with a Hamiltonian
\begin{equation}
      {\cal H}[\{\phi_i(u)\}] = \int_0^{\hbar\beta} du
                   \left[Aa\sum_{i=1}^N \frac{1}{2}\dot{\phi}_i^2(u)
                   + V(\{\phi_i(u)\})\right],
 \label{eq:4lagrangian}
\end{equation}
where the potential is given by
\begin{equation}
      V(\{\phi_i\}) = Aa \sum_{i=1}^N \left[
                \frac{\omega_0^2}{2}(\phi_i-\phi_{i-1})^2
                + \frac{\omega_1^2}{8} (\phi_i^2-1)^2 \right].
 \label{eq:4potential}
\end{equation}
Here $\beta=1/k_BT$ denotes the inverse temperature,
$\dot{\phi}_i\equiv d\phi_i/du$, and ${\cal D}[\{\phi_i(u)\}]$
is the usual path-integral measure.
The partition function describes a set of $N$ harmonically coupled
oscillators of mass $Aa$ separated by a distance $a$, with each
oscillator moving in a double-well potential.
As usual, we assume periodic boundary conditions, $\phi_0\equiv\phi_N$.

Following the notation of Ref.\cite{gtv88b} we introduce dimensionless
parameters  and define a coupling constant $Q = \hbar\om_1/E_s$
which controls the quantum 
character of the system by determining whether the kinks are ``heavy''
enough to be treated semiclassically.
In our simulations
we will fix the energy scale by setting  the energy of the classical
static kink $E_s=(2/3)Aa\om_0\om_1=1$.
We also introduce the parameter $R=\om_0/\om_1$
which measures the length of the
classical kink in units of the lattice spacing $a$.
The reduced temperature will be denoted by $t\equiv k_BT/E_s$.

%------------------------------------------------------------------
%      \section{Variational approximation}\label{sect:4VarApp}
%------------------------------------------------------------------
The variational approach for one-dimensional quantum systems
\cite{gt,gtv88a,gtv88b}
starts from a quadratic trial Hamiltonian.
The parameters in this trial Hamiltonian are determined
by optimizing the Jensen-Peierls inequality for the free energy.
A numerical solution of the resulting set of $N(N+1)/2$
self-consistent equations is extremely complicated.
Therefore only the limiting cases
of high and low temperatures and for small coupling $Q$ have been
treated in the literature. 
For the latter case, which seems to be the most useful one,
the final result reads \cite{gtv88b}
\begin{eqnarray}
\beta F &=&  \sum_{i=1}^N \ln \frac{\sinh F_k}{F_k}
        - \beta\frac{3}{4}NAa\om_1^2D^2
              \nonumber \\
     & &- \ln \Bigl\{ 
             \left[\frac{\hat{A}a}{2\pi\hbar^2\beta}\right]^{N/2}
             \prod_{i=1}^N \int d\phi_i
             \exp\Bigl[-\beta \hat{A}a \sum_{i=1}^N 
                  \left[\right.\frac{\om_0^2}{2}(\phi_i-\phi_{i-1})^2
               \nonumber \\
       & & \qquad \qquad \qquad \qquad\qquad\qquad\qquad
          +\frac{\hat{\om}_1^2}{8}(\phi_i^2-1)^2
                       \left.  \right] \Bigr]
           \Bigr\},
 \label{eq:4energy}
\end{eqnarray}
where $\hat{A}\equiv A(1-3D)$, $\hat{\om}_1^2 \equiv \om_1^2(1-3D)$,
and $D = \sum_{i=1}^N \left[\frac{\hbar^2\beta}{4AaF_k}
        \left(\coth F_k - 1/F_k\right)\right]$
with $F_k = \beta\hbar\om_k/2$ and 
$\omega_k^2 = 4\om_0^2\sin^2(k\pi/N) + \om_1^2$.

In the sequel the thermodynamic observables of interest will be the
internal energy per site 
$u = U/N = \frac{1}{N}\left(F - T\frac{\partial F}{\partial T}\right)$
and the specific heat per site $c$ given by
$c = C/N = \frac{1}{N}\frac{\partial U}{\partial T}$.
More precisely, we will be interested only in the anharmonic
contribution to these quantities. For the free energy this is given
by $dF\equiv F-F_{\rm harmon} = F-(1/\beta)\sum\ln(2\sinh F_k)$.
Therefore the last two terms
of eq.(\ref{eq:4energy}) give the anharmonic contribution to the quantum 
free energy after subtracting the corresponding classical contribution
$F_{\rm class}=(1/\beta)\sum\ln(2F_k)$.
In order to obtain analytical data for a comparison with our Monte Carlo
results we therefore have to evaluate the configurational integral
(\ref{eq:4energy}).

This can be achieved by employing standard transfer integral techniques
\cite{ss}.
Here we have to find the eigenvalues of the
transfer integral equation
associated with eq.(\ref{eq:4energy}).
In the thermodynamic limit $N\rightarrow \infty$
only the lowest eigenvalue survives
but in order to control finite-size effects we need to compute all
eigenvalues. 
With decreasing temperature more and more eigenvalues have to be taken into
account.
In particular, we observe that for low temperatures the two lowest
eigenvalues are almost degenerate.

%------------------------------------------------------------------
       \section{Simulation techniques} \label{sect:4SimTech}
%------------------------------------------------------------------

The partition function
(\ref{eq:4partitionfunction})-(\ref{eq:4potential}) 
was discretized using the Takahashi-Imada scheme \cite{ti}.
The discretized version of the partition function here reads
\begin{eqnarray}
  {\cal Z} &=& \prod_{i=1}^N \prod_{k=1}^L 
             \int \frac{d\phi_{i,k}}{\sqrt{2\pi\beta\hbar^2/LAa}}
             \exp\Bigl\{ 
                 -\frac{\beta}{L}Aa\sum_{i=1}^N\sum_{k=1}^L
                       \frac{1}{2}\left(\frac{L}{\hbar\beta}\right)^2
                       (\phi_{i,k}-\phi_{i,k-1})^2
                     \nonumber \\
            & & \qquad  \qquad  \qquad  \qquad    \qquad    \qquad      
           -\frac{\beta}{L} V_{\rm TI}(\{\phi_{i,k}\})\Bigr\},
 \label{eq:4discpartfunction}
\end{eqnarray}
where the potential is given by
\begin{equation}
   V_{\rm TI}(\{\phi_{i,k}\}) = \sum_{k=1}^L \Bigl\{ 
             V(\{\phi_{i,k}\}) + \frac{\beta^2\hbar^2}{24AaL^2}
             \sum_{i=1}^N \left(\frac{\partial V}{\partial
                      \phi_{i,k}}\right)^2\Bigr\}.
 \label{eq:4Vdiscrete}
\end{equation}
Here $k$ denotes the additional index for the Trotter discretization
at each site. The convergence of this discretization is of the order
$\epsilon^4$ where $\epsilon\equiv \hbar\beta/L$ and $L$ is the Trotter number.
The standard, quadratically convergent discretization scheme
is recovered by dropping the second term in eq.(\ref{eq:4Vdiscrete}). 

Since for local update algorithms we expect a quadratic slowing down
in the continuum limit of large Trotter numbers $L$ \cite{js93a}
we applied a multigrid W-cycle with piecewise constant interpolation
\cite{multigrid}
at each site along the Trotter direction. Note that since
we are approximating the continuum limit only for the Trotter
discretization we do not need to apply two-dimensional 
multigrid coarsening.
Also note that the interactions between the oscillators enter 
the multigrid coarsening only as constants for the polynomial
expression for the energy on the finest grid.

The observables of interest are the internal energy and the specific
heat.
As to the question of energy estimators we remark that the
discretized partition function (\ref{eq:4discpartfunction})
gives rise to a so-called kinetic estimator of the energy $U_k$ 
\cite{barker79} by differentiating
%\begin{equation}
$\langle U \rangle = - \partial \ln {\cal Z}/\partial \beta
                       \approx \overline{U_{\rm k}}$
%\end{equation}
where $\overline{U_{\rm k}}$ denotes the simple arithmetic mean
over $N_m$ measurements of $U_{\rm k}$ in the Monte Carlo process.
Applying a simple scaling argument one can find a different
but equivalent energy estimator $U_v$ \cite{virial}
based on the virial theorem with different variance.
In order to reduce the variance of the energy estimation
we may then use a linear combination of these two estimators.
In doing so it should be
noted that the optimal combination of the two estimators has to
take into account the individual variances {\em and} the
covariance of the (blocked) individual estimators \cite{jstobepub}. 
Note that the energy estimators differ for the standard
discretization scheme and the improved discretization since the 
correction term in $V_{\rm TI}$ is $\beta$-dependent.
For the evaluation of the anharmonic contributions the 
discretization error was further reduced by subtracting the
exact values for the harmonic contribution at finite Trotter number $L$.
For the standard  discretization this improvement was already made use
of in Ref.\cite{exactharmonic}. Since we are only dealing with 
Gaussian integrals the exact harmonic contribution can,
however, also readily be found for the Takahashi-Imada scheme \cite{jstobepub}.
A full account of the simulational details discussing various
systematic algorithmic refinements of path-integral Monte Carlo simulations 
will be given elsewhere \cite{jstobepub}.

       \section{Results}\label{sect:4Results}
We have performed simulations of the partition function
(\ref{eq:4partitionfunction})-(\ref{eq:4potential}) 
using the improved discretization scheme (\ref{eq:4discpartfunction}),
(\ref{eq:4Vdiscrete})
for different values of $N$,$Q$, and $t$.
The parameter $R$ was kept fixed at $R=5$ for all simulations.
The update was performed using a multigrid W-cycle with piecewise
constant interpolation in Trotter direction at each site
with single-hit Metropolis updating and $n_1=1$ pre-, $n_2=0$ postsweeps.
For each data point we have measured the internal energy
using the optimally combined estimator with
$N_m=200000$ measurements taken every second sweep, i.e. $n_e=2$,
after discarding $n_e\times 1000$ sweeps for thermalization.
The Metropolis acceptance rates were adjusted to be $\approx 40-60\%$
on the finest grid and the same step width was used for all
multigrid levels.
The specific heat was measured by simple numerical differentiation
of the ``combined'' estimator which was reweighted 
in a temperature interval of $dt=0.0001$.
These estimates gave consistent values with direct measurements
of the specific heat using the estimators obtained by differentiating
the discrete partition function but (slightly) smaller errors.
All statistical errors were computed by jackkniving the data on the basis
of $500$ blocks.

Comparing the jackknife error with the canonical variance of
the individual measurements we find that the
integrated autocorrelation time for both the kinetic energy estimator
and the virial estimator never exceeded a value of
$\tau^{\rm int}/n_e\leq 2$. Within these bounds 
we noticed that the autocorrelation times tended to be larger for low
temperatures and large coupling constants.
This observation is also reflected in the fact that the acceptance
rates were roughly constant on all levels for high temperatures
and small couplings but tended to decrease for lower $t$ and
larger $Q$.
We conclude that in our simulations the measurements of
the energies were more or less statistically decorrelated.

\begin{figure}[tb]
\vskip  10.0 truecm
\includegraphics{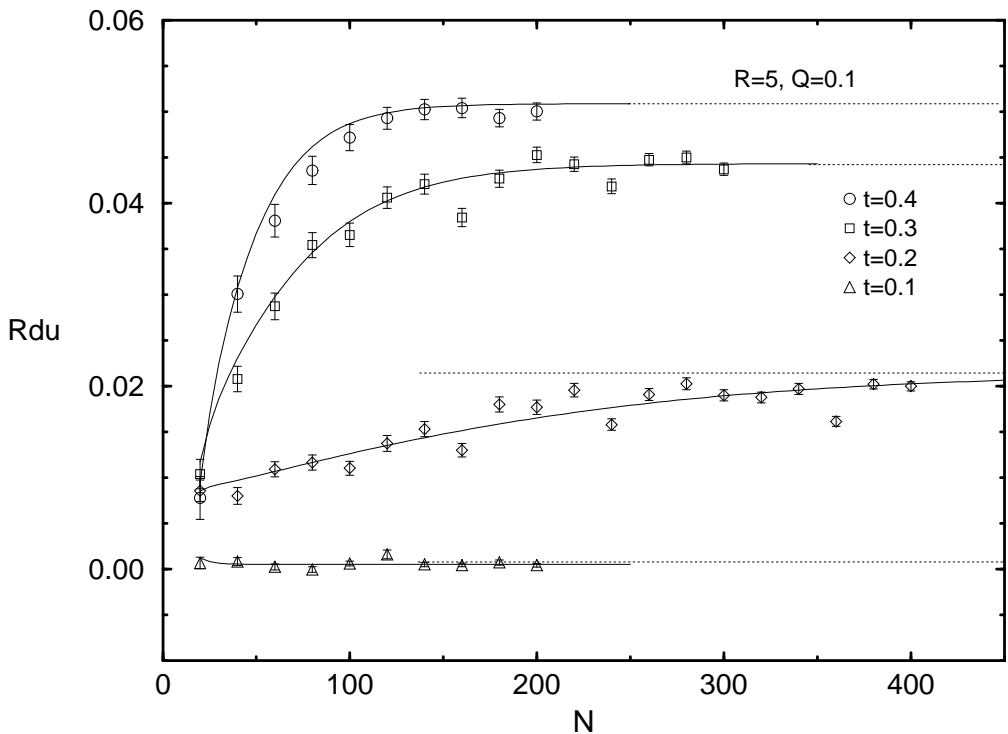}
\caption[Anharmonic contributions to the internal energy $Rdu$ as
a function of particle number $N$ for different temperatures]{%
Finite-size dependence for the 
measured anharmonic contributions to the internal energy per site.
Solid lines show the variational approximation for finite $N$
and dotted lines show the thermodynamic limit.
}
\label{figure:4nt}
\end{figure}

Figure \ref{figure:4nt} shows the measured anharmonic contributions
to the internal energy per site 
for $Q=0.1$ and $t=0.1,0.2,0.3$, and $0.4$ as
a function of the number of oscillators $N$.
Here the Trotter number $L$ was set to $L=16$ for all temperatures.
The solid lines show the variational approximation
for finite $N$, and the
dotted horizontal lines show the corresponding values
in the thermodynamic limit $N\rightarrow\infty$.
We see that the Monte Carlo data fully confirm the variational
approximation within the statistical uncertainty.
For high temperatures the finite-size effects are quite appreciable
but die off rapidly with increasing $N$.
For low temperatures on the other hand
the finite-size data approach the thermodynamic limit
rather slowly but the absolute values differ only by a small
amount from the asymptotic value.

\begin{figure}[htb]
\vskip  10.0 truecm
\includegraphics{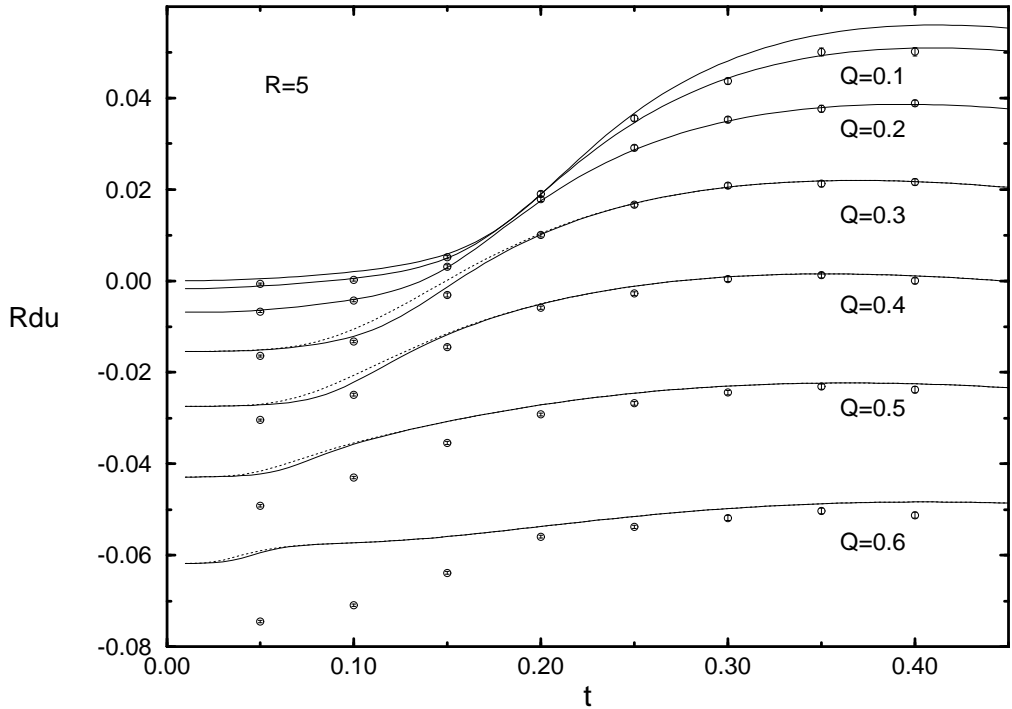}
\caption[Anharmonic contributions to the internal energy $Rdu$ as
a function of temperature $t$ for various couplings $Q$]{%
Anharmonic contributions to the internal energy per site as a function
of the temperature $t$ for various coupling parameters $Q$. 
Solid lines show the variational approximation for $N=300$.
Dotted lines show the variational approximation for $N=\infty$.
}
\label{figure:Rdu}
\end{figure}

Let us now look at the temperature dependence of the internal energy.
Figure~\ref{figure:Rdu} shows the measured anharmonic contributions
to the internal energy per site as a function of the temperature $t$ for
various couplings $Q$.
Here the number of oscillators was $N=300$ except for $t=0.05$, $0.30$,
$0.35$, and $0.40$
where we simulated a chain of $N=200$ oscillators. The Trotter 
number was $L=16$ for $t\geq 0.20$, $L=32$ for $t=0.15$, $L=64$ for $t=0.10$, 
and $L=128$ for $t=0.05$.
Regarding a comparison with the variational data we observe that
the approximation again is fully confirmed for high temperatures $t$
and small couplings $Q$.
For lower $t$ we still find a satisfactory agreement 
if we also take into account finite-size corrections.
The situation is different, however, for low temperatures 
and large couplings as can be clearly seen in Fig.~\ref{figure:Rdu}.
Here we find significant deviations from the variational approximation.
Note that the error bars for the data are
smaller than the data symbols.
Let us take a closer look at the lowest temperature which we
have investigated, $t=0.05$. 
For $Q=0.1$ and $Q=0.2$ the variational approximation is confirmed
within our statistical error estimate. But already for $Q=0.3$ we find 
a statistically significant discrepancy between our measured value
of $Rdu=0.01641(21)$ and the variational approximation which predicts
a value of $Rdu=-0.01512$ for both $N=200$ and $N=\infty$.
This discrepancy increases if we go to larger couplings. For the worst
case, $Q=0.6$,
the variational approximation gives a value of $Rdu=-0.05898$ for
$N=\infty$ and $Rdu=-0.05982$ for $N=200$. The simulation on the other
hand yields a value of $Rdu=-0.07447(26)$,
%(cp. Table \ref{table:4Rdu}),
i.e., the variational
approximation deviates from the Monte Carlo results by $56$ 
statistical error bars even if finite-size corrections are fully taken
into account.

In order to check whether for the Monte Carlo data systematic errors
due to the discretization
would still play a role we have checked our data for $t=0.1$ against
simulations with smaller Trotter numbers $L=16$ and $L=32$.
For $Q=0.1$ and $Q=0.2$
we found no significant differences but we did observe finite $\eps$
effects for larger couplings $Q$. Their size, however, was 
small enough and in view of the fact that our discretization converges
with the fourth order in $\eps$ we believe that the remaining 
discretization error for small $t$ and large $Q$ is at most of
the same order as the statistical errors. In any case, we observed that
going to larger Trotter number would push the values { \em down}, i.e. would
further increase the difference to the variational approximation.

We conclude that our data differ significantly from the variational
approximation for large $Q$ and small $t$.
The question then arises whether these discrepancies are a consequence
of the low coupling expansion or rather inherent to the variational 
approach at this level. In view of the fact that the data fit
quite well even for large $Q$ at high temperatures
it seems more likely that the latter is the case. 
On the basis of the validity of the Wigner expansion Giachetti {\em et al}.
\cite{gtv88b}
suggested that their expansion be valid as long as
\begin{equation}
     t \gg \frac{1}{8\pi}Q^2\ln(8R)\approx 0.1468 Q^2.
 \label{eq:valid}
\end{equation} 
This means, for $Q=0.3$ we have to compare $t=0.05$ with $0.013212$ 
to explain a discrepancy of almost $5$ statistical error bars.
For our worst case of $Q=0.6$ eq.(\ref{eq:valid}) reads explicitly
$t\gg 0.053$ and the
discrepancy of $56$ statistical error bars for $t=0.05$ is indeed due to a
violation of this condition. Looking at Fig.~\ref{figure:Rdu}
for the largest coupling, $Q=0.6$, we conclude
that condition (\ref{eq:valid}) is violated for almost all temperatures
displayed in Fig.~\ref{figure:Rdu} only if we take the ``$\gg$''
to mean: ``larger by more than one order of magnitude.''

\begin{figure}[tb]
\vskip  10.0 truecm
\includegraphics{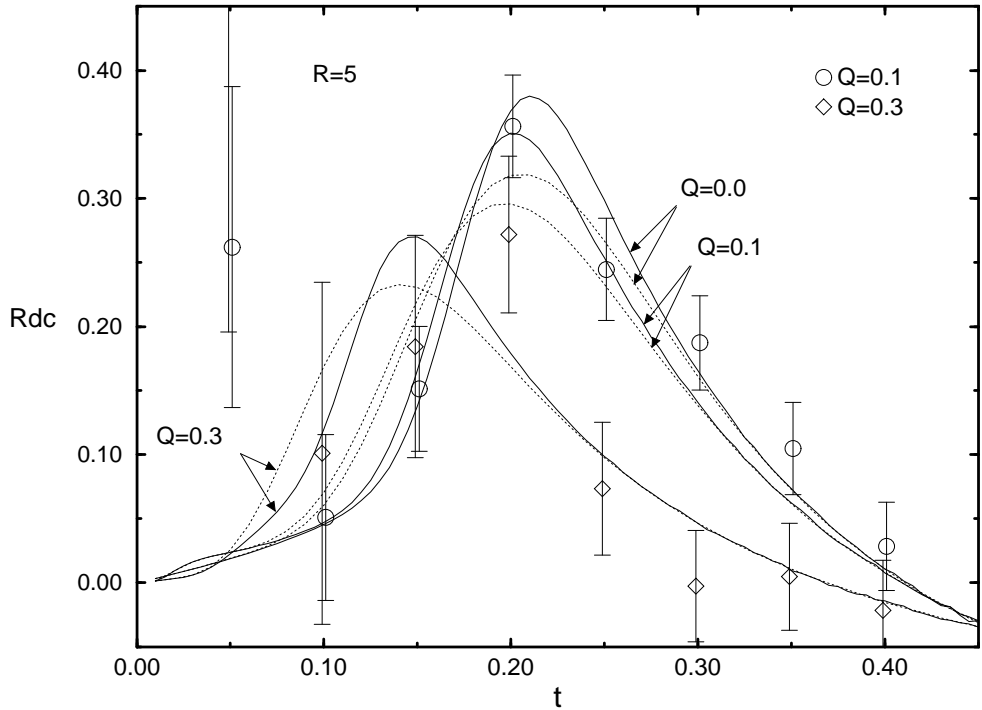}
\caption[Anharmonic contributions to the specific heat $Rdc$ as
a function of temperature $t$ for different couplings $Q$]{%
Anharmonic contributions to the specific heat per site as a function
of the temperature $t$ for various coupling parameters $Q$. 
Solid lines show the variational approximation for $N=300$.
Dotted lines show the variational approximation for $N=\infty$.
}
\label{figure:Rdc}
\end{figure}

Let us finally take a look at the specific heat. 
Figure \ref{figure:Rdc} shows the measured anharmonic contributions
to the specific heat per site.
Again the solid lines are the variational approximation for $N=300$
and the dotted lines show the corresponding thermodynamic limit.
Due to the fact that the estimation of the specific heat involves
a difference of statistically fluctuating variables the resulting
statistical accuracy is greatly reduced compared to the
estimation of energies. Therefore our data for the specific heat 
do not allow for
a significant falsifying test of the variational approximation.
For the more interesting case of large couplings we also see
that the statistical uncertainty unfortunately is even increasing,
in particular for low temperatures.
One therefore would have to conclude that neither a brute force
increase of the statistics appears to be
a reasonable way of getting more accurate data.
With these restrictions we nevertheless do see, however, that the
general trend of the quantum effects
as computed by the variational approximation is confirmed.

%------------------------------------------------------------------
       \section{Discussion}\label{sect:4Discussion}
%------------------------------------------------------------------

Employing refined path-integral Monte Carlo techniques we have
been able to considerably reduce the 
systematic and statistical errors of a quantum
Monte Carlo simulation of the $\phi^4$ chain.
The resulting accuracy now allows for a significant test of the
variational approximation. For small couplings we find that the
variational quantum corrections to the thermodynamic quantities
are fully confirmed and only for large couplings and low temperatures
do we observe significant deviations from the exact Monte Carlo data.
The discrepancies may be due to the additional approximation
of the small coupling expansion which was used to evaluate the 
effective classical potential of the variational approximation.
It would therefore be interesting to see whether the Monte Carlo data
might be reproduced by taking into account higher-order corrections
in the coupling parameter $Q$.
The discrepancies increase both for large couplings and for low
temperatures. Since it is known that the variational approximation
works better at high temperatures the 
other possible reason for the deviations of the analytical
data may be an inherent failure of the variational
approximation itself (at this level of accuracy). If this should be the
case it would be interesting
to see whether by calculating the higher-order corrections to the
variational approach \cite{kleinert93a} one would be able to
account for the remaining discrepancies.

\section*{Acknowledgments}
We wish to thank R.~Vaia for helpful comments.
W.J. thanks the Deutsche Forschungsgemeinschaft for a Heisenberg
fellowship.
%-----------------------------------------------------------------------
               
% 	       	           
\end{document}